\documentclass[a4paper]{ifacconf}

\usepackage{graphicx,amsmath,url}      
\usepackage{amsmath,amssymb,amsfonts}
\usepackage[round]{natbib}             

\def\portugues{1} 

\ifx\portugues\undefined
\def\portugues{0}
\fi

\if\portugues0
   \usepackage[english]{babel}
  \else
   \usepackage[spanish,brazil,english]{babel}
\fi

\usepackage[T1]{fontenc}

\usepackage[utf8]{inputenc}

\usepackage{ae}

\if\portugues1
\usepackage{subfigure}

\newtheorem{defin}{Defini\c{c}\~ao}
\newtheorem{theo}{Theorem}
 { }
 { }
 { }
 
\fi

\begin{document}

\if\portugues1

\begin{frontmatter}

\title{Some remarks on the performance of Matlab, Python and Octave in simulating dynamical systems} 

\thanks[footnoteinfo]{This work has been supported by the Brazilian agency CAPES.}

\author[First]{Priscila F. S. Guedes,} 
\author[Second]{Erivelton G. Nepomuceno} 

\address[First]{Graduate Program in Electrical Engineering - Universidade Federal de Minas Gerais - Av. Antônio Carlos 6627, 31270-901, Belo Horizonte, MG, Brazil (e-mail: pri12\_guedes@hotmail.com).}
\address[Second]{Control and Modelling Group (GCOM), Department of Electrical Engineering, Federal University of São João del-Rei, São João del-Rei, MG, 36307-352, Brazil
b (e-mail: nepomuceno@ufsj.edu.br)}
   
\renewcommand{\abstractname}{{\bf Abstract:~}}   
   
\begin{abstract}                
Matlab has been considered as a leader computational platform for many engineering fields. Well documented and reliable, Matlab presents as a great advantage its ability to increase the user productivity. However, Python and Octave are among some of the languages that have challenged Matlab. Octave and Python are well known examples of high-level scripting languages, with a great advantage of being open source software. The novelty of this paper is devoted to offer a comparison among these tree languages in the simulation of dynamical systems. We have applied the lower bound error to estimate the error of simulation. The comparison was performed with the chaotic systems Duffing-Ueda oscillator and the Chua's circuit, both identified with polynomial NARMAX. Octave presents the best reliable outcome. Nevertheless, Matlab needs the lowest time to undertake the same activity. Python has presented the worse result for the stop simulation criterion.
\end{abstract}


\end{frontmatter}
\fi


\section{Introduction}
\selectlanguage{english}
\hyphenation{FAPEMIG NARMAX}

Computational simulations are fundamental in analysis of nonlinear dynamical systems \cite{Gal2013,Loz2013,SGY1997,HYG1987}. Among the used softwares, Matlab stands out due to its high performance oriented to the numerical calculation. Additionally, Matlab has been used to develop many well-cited toolbox to control theory, such as YALMIP \cite{lofberg2004yalmip} or System Identification Toolbox \cite{ljung2009developments}.  Octave is free software that is more compatible with Matlab. And in recent years, Python software has become popular and its popularity is also ranked first in the IEEE Spectrum ranking for 2018 \cite{IEEEspectrum}. The three software are examples of scripting languages, that is, these languages are interpreted.

These softwares are used in digital computers that have inherent properties  which numerical simulations do not present exact results  \cite{Gal2013}. This fact occurs due to the representation of the real number on computer, which may cause approximations, rounding and truncation errors \cite{Lia2009}. Error propagation control is considered highly important, especially when they characterize chaotic systems, since small errors introduced in each computational step can grow exponentially due to the high sensitivity presented in chaotic systems \cite{MN2016,NM2016}. In addition to rounding errors, truncation errors are introduced during integration of continuous time systems by using numerical methods, which are constructed by skipping higher-order terms in the Taylor expansion of the solution, these errors decrease computational reliability \cite{QL2017}. 

\cite{Loz2013} states that there are several published works related to chaotic dynamical systems, which the results were not carefully checked, compromising the reliability of the results. In this context,  \cite{Nep2014} showed that the iterations of the Logistic Map generated a result that mathematically was not what was expected. In addition, the use of different discretization methods results in different numerical solutions
 \cite{Lia2009,NM2017}.

In order to investigate numerical error in system simulations,  \cite{NMAR2017} proposed a method to calculate the Lower Bound Error (LBE) based on the fact that two equivalent mathematical extensions can generate divergent results on computational simulations. Based on this method there was an expansion to an arbitrary number of mathematical extensions \cite{GPB+2017,CNG+2006,Unpingco2008,FC2017}. In this sense, few studies have considered the influence of the programming language on the reliability of numerical solutions
\cite{JBN+2017}.  Usually the comparison of such languages has been devoted to time consumption or arithmetic and algebraic operations, as done by \cite{4755916}. The novelty of this paper is devoted to offer a comparison among these tree languages in the simulation of dynamical systems.

The rest of the paper is organized as follows. In Section \ref{sec:cp} we recall some preliminary concepts. Then, in Section \ref{sec:met}, we present the developed method. Section 4 is devoted to present the results, then the final remarks are given in Section 5. 

 \section{Preliminary concepts}
 \label{sec:cp}
 
 \subsection{The polynomial NARMAX}

The NARMAX model is a representation for nonlinear systems.
This model can be represented as  \cite{CB1989} 
\begin{eqnarray}
 y(k) &=& F^l[y(k-1),\cdots ,y(k-n_y),  \nonumber\\
 &&u(k-1),  \cdots , u(k-n_u),   \\ 
&&e(k-1), \cdots , e(k-n_e)]+ e(k), \nonumber 
\end{eqnarray}
where $y(k)$, $u(k)$ e $e(k)$ are, respectively, the output, the input and the noise terms at the discrete time $n \in \mathbb{N}$. The parameters $n_y$, $n_u$ e $n_e$ are their maximum delay. And $F^{\ell}$ is a nonlinear function of degree $\ell$.

\subsection{Recursive functions}

In recursive functions is possible to calculate the state $x_{n+1}$, at a give time, from an earlier state $x_n$
\begin{equation}
   x_{n+1}=f(x_n), 
\end{equation}
where $f$ is a recursive function and $x_n$ is a function state at the discrete time $n$. Given an initial condition $x_0$ and with successive applications of the function $f$ it is possible to know the sequence $x_n$ \cite{gilmore2012topology}.

\subsection{Natural interval extension}
The natural interval extension is achieved by changing the sequence of arithmetic operation \cite{moore2009}, that is, the extensions are mathematically equivalents.

Furthermore, two extension which algebraically are the same function may not be equivalent in interval arithmetic

\textbf{Example 2.3}: Based on the logistic map \cite{May1976}, interval extensions are:

\begin{eqnarray}
 x_{n+1}=rx_n(1-x_n)  \nonumber\\
 x_{n+1}=rx_n - rx^2_n \nonumber \\
 x_{n+1}=rx_n - rx_nx_n \nonumber
\end{eqnarray}

\subsection{Orbits and pseudo-orbits}
Associated with a map we may define an orbit as follows \cite{HYG1987}:
\begin{defin}
\label{def3}
The true orbit $\{x_n\}^N_{n=0}$ satisfies $x_{n+1}=f(x_n)$.
\end{defin} 
That is, given an initial condition $x_0$, and interacting the function, a sequence of values represented by $\{x_n\}=[x_0,x_1,\cdots,x_n]$ is defined. When the computer is used to calculate the recursive functions, numeric errors are propagated during successive calculations, then the true orbit is not calculated but a representation of the same, which is called pseudo-orbit

\begin{eqnarray}
\{\hat{x}_{i,n}\}=[\hat{x}_{i,0},\hat{x}_{i,1},\cdots,\hat{x}_{i,n}],
\end{eqnarray}
which accepts the relation
\begin{equation}
\label{eq:orb.1}
    |x_n-\hat{x}_{i,n}|\le\delta_{i,n},
\end{equation}
where $\delta_{i,n} \in \mathbb{R}^+$.

Thus, we define an interval associated with each value of a pseudo-orbit
\begin{equation}
\label{eq:orb.2}
    I_{i,n}=[\hat{x}_{i,n}-\delta_{i,n}\,,\,\hat{x}_{i,n}+\delta_{i,n}].
\end{equation}
From Equations (\ref{eq:orb.1}) and (\ref{eq:orb.2}) it is clear that
\begin{equation}
    x_n \in I_{i,n} \quad\mbox{for all i} \in \mathbb{N}.
\end{equation}

\subsection{The lower bound error}

The  lower  bound  error  consists  of  a  tool  to analyze the error propagation in numerical simulations proposed by \cite{NMAR2017}.

\begin{theo}
\label{teo}
Let two pseudo-orbits $\{\hat{x}_{a,n}\}$ and $\{\hat{x}_{b,n}\}$ derived from two natural interval extensions. Let $\ell_{\Omega,n}=| \hat{x}_{a,n} - \hat{x}_{b,n}|/2$ be the lower bound error associated to the set of pseudo-orbits $\Omega=[\{\hat{x}_{a,n}\},\{\hat{x}_{b,n}\}]$ of a map, then $\gamma_{a,n}=\gamma_{b,n} \ge \ell_{\Omega,n}$.
\end{theo}

The proof of this theorem can be found in \cite{NMAR2017}.

\section{Methods}
\label{sec:met}

\cite{NMAR2017} developed the Lower Bound Error theorem. But, different software can cause different result. Hence, the objective is to investigate and compare the Matlab R2016a-64 bits, Octave 4.2.1 - 64 bits and Python 3.5.4-64 bits performances using the Theorem \ref{teo} and the polynomial NARMAX for the Duffing-Ueda oscillator and Chua's circuit.

The proposed method can be summarized in the following steps:
\begin{enumerate}
    \item \textbf{Step 1}: choose two natural interval extensions for each system;
    \item \textbf{Step 2}: calculate the system's orbit from the chosen extensions in each software;
    \item \textbf{Step 3}: determine the lower bound error;
    \item \textbf{Step 4}: compare the results obtained in each software. This parallel will be performed using a stop simulation criterion and verifying in how many iterations each software reached that criterion. And also by the time each software spends to simulate the algorithms.
\end{enumerate}

\subsection{Stop simulation criterion}

It was used a method which verifies the loss of simulation accuracy as a stop criterion. Then, $\varepsilon_{\alpha,n}$ is the relative precision at iteration $n$ defined by

\begin{equation}
    \varepsilon_{\alpha,n} = \frac{\hat{x}_{a,n}-\hat{x}_{b,n}}{\hat{x}_{a,n}+\hat{x}_{b,n}}
    \label{eq:stop}
\end{equation}

\noindent where $n \in \mathbb{N} $, $\hat{x}_{a,n}$ and $\hat{x}_{b,n}$ are the two chosen pseudo-orbits $a$ and $b$. A minimum precision, $\varepsilon$, is defined, which implies that the simulation would be stopped at the moment when $\varepsilon_{\alpha,n} > \varepsilon$. In this work, we adopt $\varepsilon = 0.001$.

To perform the tests, it was used a computer with a processor Intel Core i5-3317U @ 1.7GHz and a Windows 10 Home Single Language operating system. All data, routines and simulations used in this work are available upon request.

 \section{Numerical experiments}

\subsection{Duffing-Ueda}

Considering a damped, periodically forced nonlinear Duffing-Ueda oscillator \cite{Billings2013}:
\begin{equation}
    \frac{d^2y}{dt^2}+k\frac{dy}{dt}+ \mu y^3=A\cos(t).
\end{equation}
where $\mu$ is the cubic stiffness parameter, $k$ is a linear damping and $A$ is the amplitude of excitation. A polynomial NARMAX for the Duffing-Ueda oscillator was identified by \cite{AB1994}.
\begin{eqnarray}
\mathrm{y_{n+1}}&=& \mathrm{2.1579y_n-1.3203y_{n-1}+0.16239y_{n-2}} \nonumber\\
&& \mathrm{+0.0003416u_n +0.001963u_{n-1}} \nonumber\\
&& \mathrm{-0.0048196y^3_n+0.003523y^2_ny_{n-1}} \label{eq:duffing}\\ &&  \mathrm{-0.0012162y_ny_{n-1}y_{n-2}+0.0002248y^3_{n-2}\nonumber }  
\end{eqnarray}
\noindent where $u=A \cos(kT_s)$, $n \in \mathbb{N}$ and $T_s= \pi /60$.

Let us consider two interval extensions of the model \ref{eq:duffing}:
\begin{eqnarray}
\mathrm{\scriptstyle F(X_n)}&=& \scriptstyle \mathrm{2.1579X_n-1.3203X_{n-1}+0.16239X_{n-2}} \nonumber\\
&&\scriptstyle \mathrm{+0.0003416U_n +0.001963U_{n-1}} \nonumber\\
&&\scriptstyle \mathrm{\underline{-0.0048196X^3_n}+0.003523X^2_nX_{n-1}} \label{eq:duffing1}\\ &&\scriptstyle \mathrm{-0.0012162X_nX_{n-1}X_{n-2}+0.0002248X^3_{n-2}\nonumber }  
\end{eqnarray}
\begin{eqnarray}
\mathrm{\scriptstyle G(X_n)}&=& \scriptstyle \mathrm{2.1579X_n-1.3203X_{n-1}+0.16239X_{n-2}} \nonumber\\
&&\scriptstyle \mathrm{+0.0003416U_n +0.001963U_{n-1}} \nonumber\\
&&\scriptstyle \mathrm{\underline{-0.0048196X^2_nX_n}+0.003523X^2_nX_{n-1}} \label{eq:duffing2}\\ &&\scriptstyle \mathrm{-0.0012162X_nX_{n-1}X_{n-2}+0.0002248X^3_{n-2}\nonumber }  
\end{eqnarray}

\begin{figure}[ht!]
\centering
\subfigure[Pseudo-orbits obtained from Python.]{
\includegraphics[width=8cm,height=6cm]{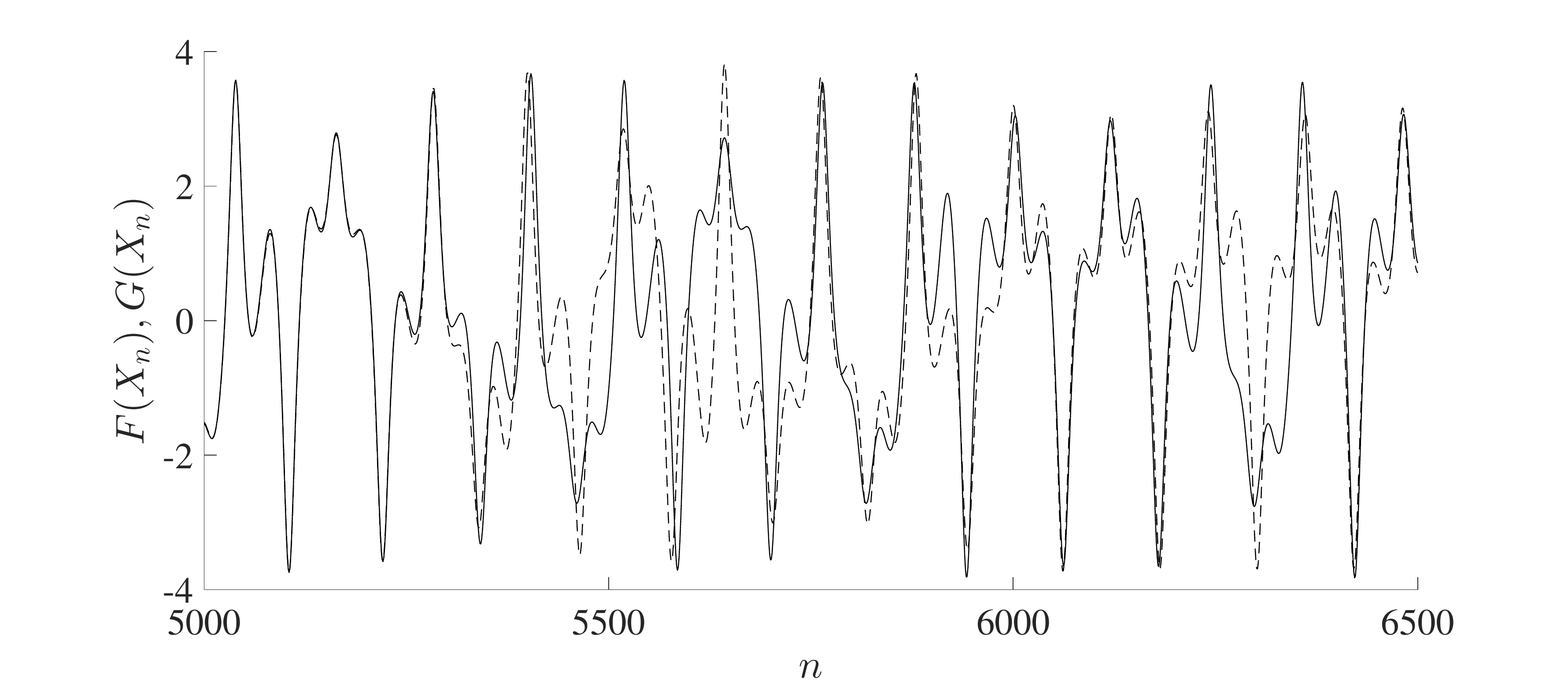}} \label{fig:1a}
\subfigure[Pseudo-orbits obtained from Matlab.]{
\includegraphics[width=8cm,height=6cm]{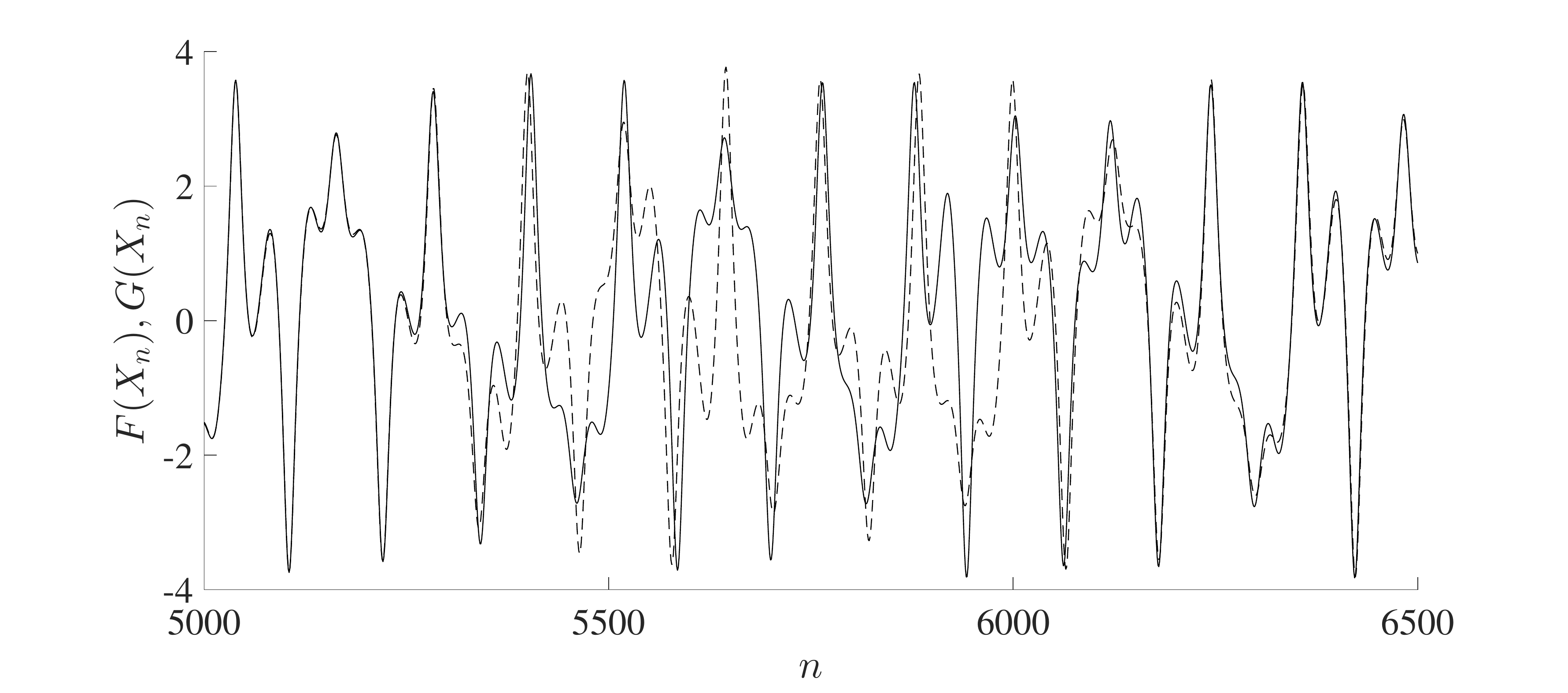}} \label{fig:1b}
\subfigure[Pseudo-orbits obtained from Octave.]{
\includegraphics[width=8cm,height=6cm]{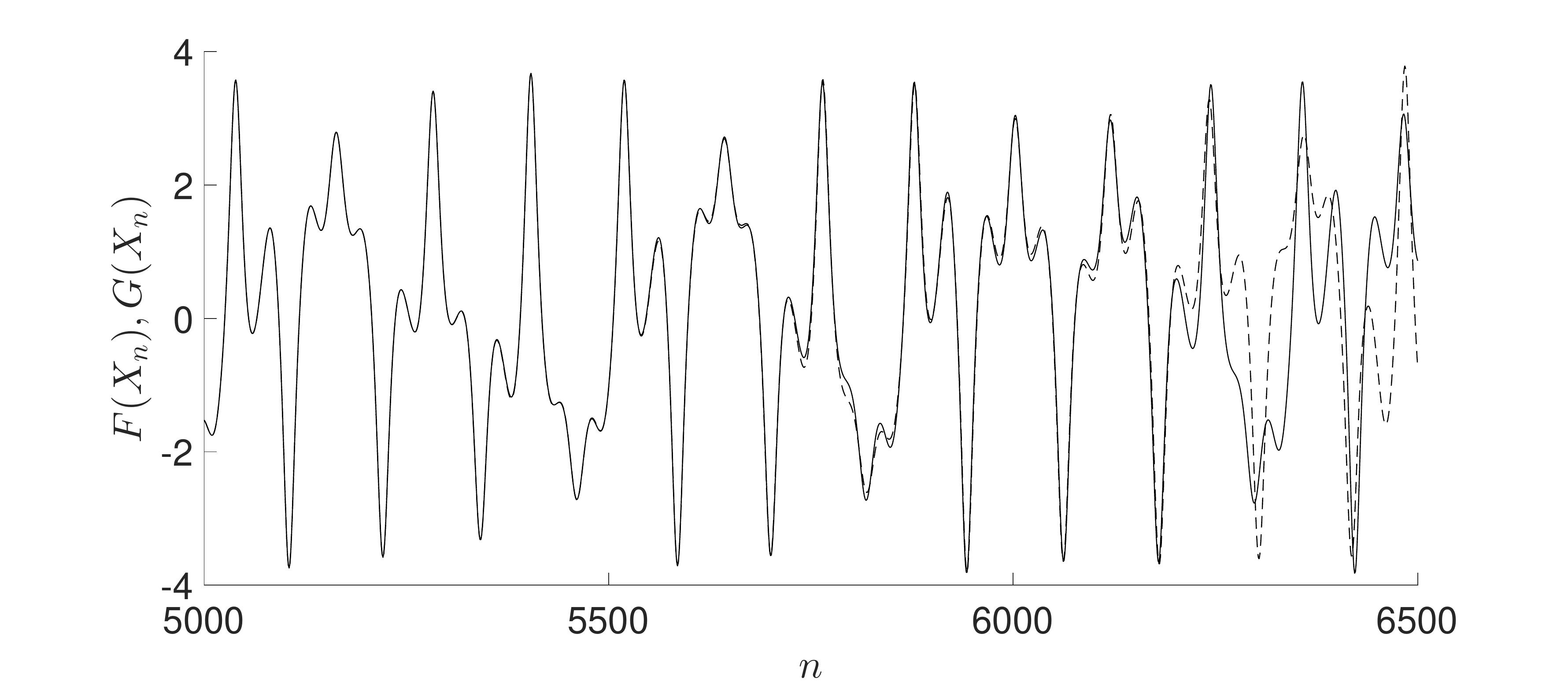}} \label{fig:1c}
\caption{ Duffing-Ueda Oscillator: Free-run simulation for the interval extensions of Equations \eqref{eq:duffing1} and \eqref{eq:duffing2}, with results for $F(X_n) (-)$ and $G(X_n) (--)$ and $n$ stands for the number of iterations.}
\label{fig:1}
\end{figure}

\begin{figure}[ht!]
\centering
\subfigure[LBE obtained from Python.]{
\includegraphics[width=8cm,height=6cm]{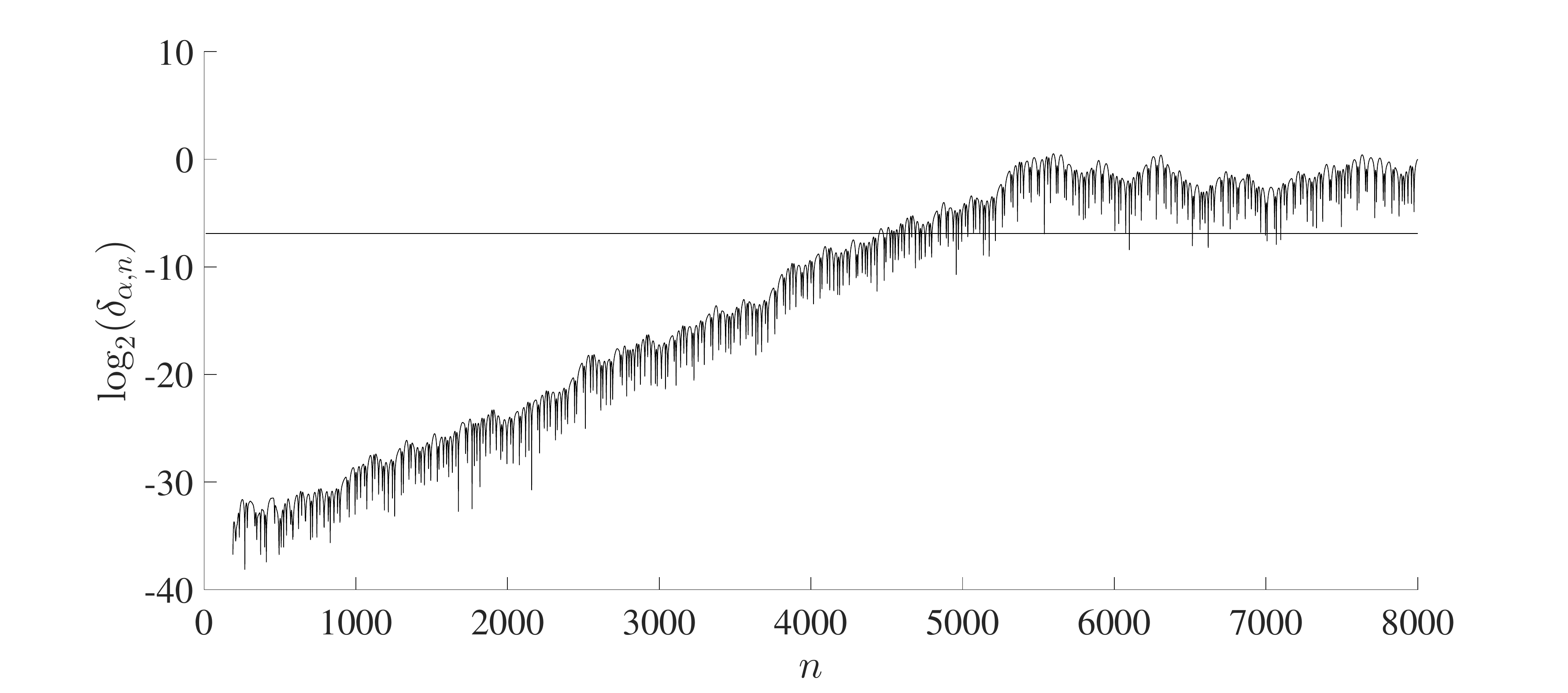}} \label{fig:2a}
\subfigure[LBE obtained from Matlab.]{
\includegraphics[width=8cm,height=6cm]{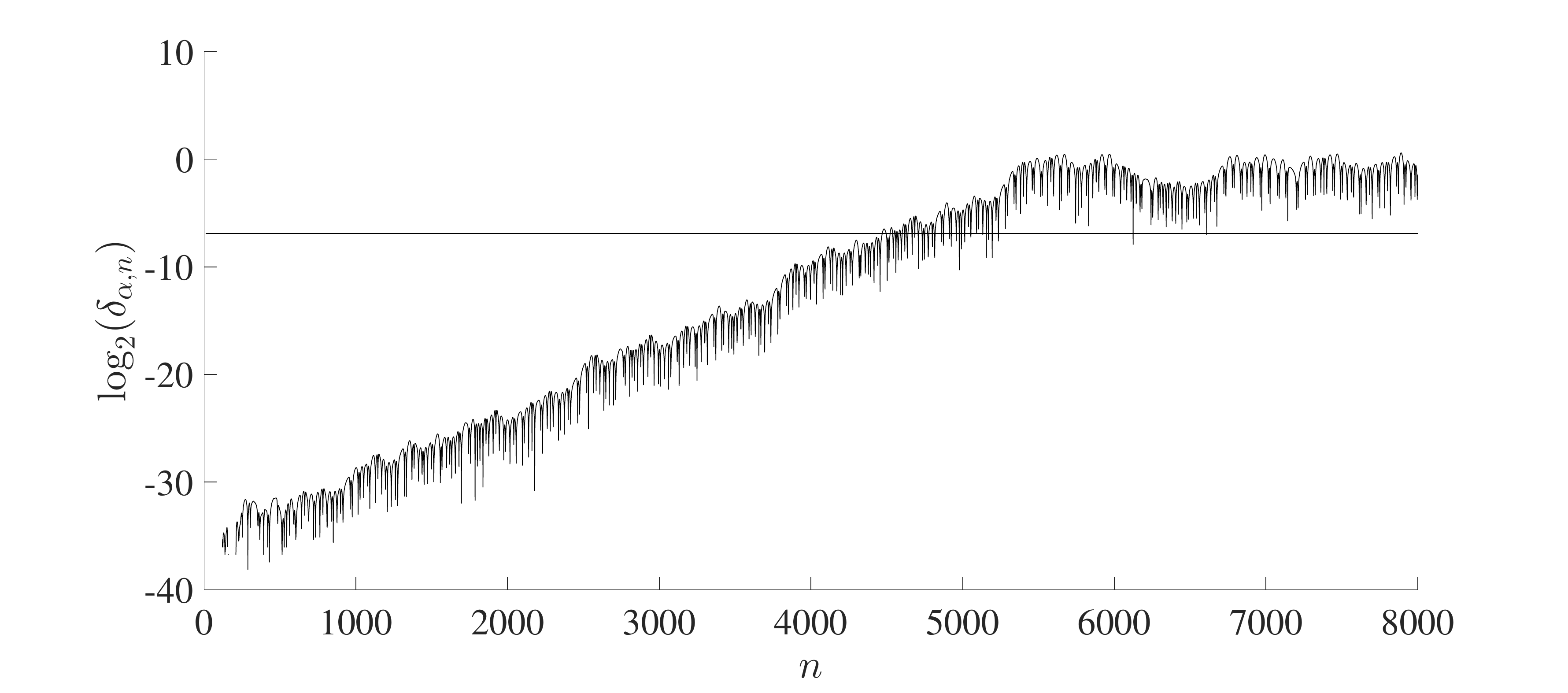}} \label{fig:2b}
\subfigure[LBE obtained from Octave.]{
\includegraphics[width=8cm,height=6cm]{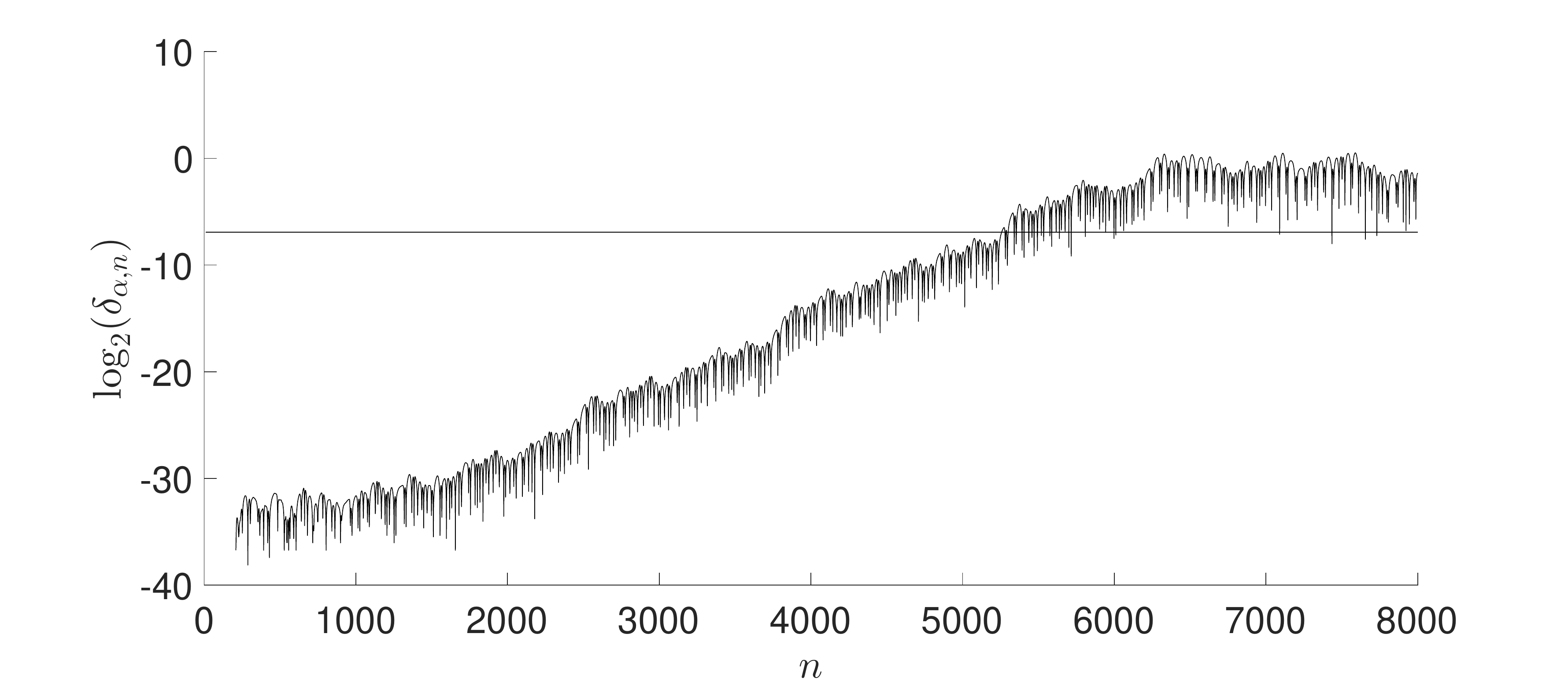}} \label{fig:2c}
\caption{ Duffing-Ueda Oscillator: Evolution of Lower Bound Error $\ell_{\Omega,n}$. The values are plotted using $log_2$.}
\label{fig:2}
\end{figure}

\noindent Figure \ref{fig:1} shows the free-run simulation for Duffing-Ueda oscillator. For this system, it can be observed that running different software cause a different pseudo-orbit. Figure \ref{fig:2} shows the evolution of the Lower Bound Error. Analyzing the number of iterations that satisfies $\varepsilon = 0.001$, the simulation is no longer reliable when $n \ge 4450$ for Python, $n \ge 4471$
for Matlab and $n \ge 5258$ for Octave. Therefore, Matlab represents a $0.47\%$ greater confidence compared to Python and Octave a $18.16\%$ greater confidence to Python.

\subsection{Chua's circuit}
\begin{figure}[h]
	\centering
	\includegraphics[width=0.9\linewidth]{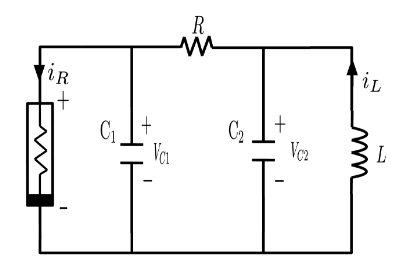}
	\caption{Chua's circuit.}
	\label{chua}
\end{figure}

The Chua's circuit (see Figure \ref{chua}) \cite{chua1993} is composed of passive linear elements (two capacitors, one inductor and one resistor) connected to an active nonlinear component, known as the Chua's diode. The Chua's circuit is able to reproduce different regimes, such as periodic and chaotic oscillations. Its equations are described as follows.

\begin{equation}
\left\{ \begin{array}{rcl}
C_1 \cfrac{dv_{c_1}}{dt} &=&\cfrac{v_{c_2} - v_{c_1}}{R} - i_R(v_{c_1})\\ C_2 \cfrac{dv_{c_2}}{dt} &=&\cfrac{v_{c_1} - v_{c_2}}{R} + i_L\\
L \cfrac{di_L}{dt} &=& - v_{c_2} 
\end{array}\right.
\label{eq.Chua}
\end{equation}

The current through the nonlinear element, $ i_R (v_{C_1}) $ is given by equation \eqref{eq.DiodoChua}:

\begin{equation}
i_R(v_{c_1}) = \begin{cases}
m_0\upsilon_1+B_p(m_0-m_1)  & v_{c_1} < -B_p,\\
m_1\upsilon_1 & |v_{c_1}| \leq B_p, \\
m_0\upsilon_1+B_p(m_1-m_0)  & v_{c_1} > -B_p,
\end{cases} 
\label{eq.DiodoChua}
\end{equation}

\noindent where $m_0$, $m_1$ and $B_p$ are the slopes and the breaking points of the nonlinear element, respectively.

A polynomial NARMAX identified for Chua's circuit is given by \cite{aguirre1997}.

\begin{small}
\begin{eqnarray}
\mathrm{y_{n+1}}&=& \mathrm{3.523y_n-4.2897y_{n-1}-0.2588y_{n-3}} \nonumber\\
&& \mathrm{-1.7784y_n^3+2.0652y_{n-2}+6.1761y_n^2y_{n-1}} \nonumber\\
&& \mathrm{+0.1623y_ny_{n-1}y_{n-3}-2.7381y_n^2y_{n-2}} \label{eq:chua}\\ &&\mathrm{-5.5369y_ny_{n-1}^2+0.1031y_{n-1}^3+0.4623y_{n-3}^3} \nonumber \\
&&\mathrm{-0.5247y_{n-1}^2y_{n-3}-1.8965y_ny_{n-2}^2} \nonumber \\
&&\mathrm{+5.4255y_ny_{n-1}y_{n-2}+0.7258y_{n-1}y_{n-3}^2} \nonumber \\
&&\mathrm{-1.7684y_{n-2}y_{n-3}^2+1.1800y_{n-2}^2y_{n-3}, \nonumber} 
\end{eqnarray}
\end{small}
\noindent where the time interval between $n$ and $n+1$ is $12$ $\mu$ s.

Considering two natural interval extensions of model \ref{eq:chua}, we have

\begin{eqnarray}
\scriptstyle \mathrm{F(X_n)}&=& \scriptstyle \mathrm{3.523X_n-4.2897X_{n-1}-0.2588X_{n-3}} \nonumber\\
&& \scriptstyle \mathrm{-1.7784X_n^3+2.0652X_{n-2}+\underline{6.1761X_n^2X_{n-1}}} \nonumber\\
&&\scriptstyle \mathrm{+0.1623X_nX_{n-1}X_{n-3}-2.7381X_n^2X_{n-2}} \nonumber \label{eq:chua1}\\ &&\scriptstyle \mathrm{-5.5369X_nX_{n-1}^2+0.1031X_{n-1}^3+0.4623X_{n-3}^3} \nonumber \\
&&\scriptstyle \mathrm{-0.5247X_{n-1}^2X_{n-3}-1.8965X_nX_{n-2}^2}  \\
&&\scriptstyle \mathrm{+5.4255X_nX_{n-1}X_{n-2}+0.7258X_{n-1}X_{n-3}^2} \nonumber \\
&&\scriptstyle \mathrm{-1.7684X_{n-2}X_{n-3}^2+1.1800X_{n-2}^2X_{n-3}, \nonumber} 
\end{eqnarray}
\begin{eqnarray}
\scriptstyle \mathrm{G(X_n)}&=& \scriptstyle \mathrm{3.523X_n-4.2897X_{n-1}-0.2588X_{n-3}} \nonumber\\
&&\scriptstyle \mathrm{-1.7784X_n^3+2.0652X_{n-2}+\underline{6.1761X_nX_nX_{n-1}}} \nonumber\\
&&\scriptstyle \mathrm{+0.1623X_nX_{n-1}X_{n-3}-2.7381X_n^2X_{n-2}} \nonumber \label{eq:chua2}\\ &&\scriptstyle \mathrm{-5.5369X_nX_{n-1}^2+0.1031X_{n-1}^3+0.4623X_{n-3}^3} \nonumber \\
&&\scriptstyle \mathrm{-0.5247X_{n-1}^2X_{n-3}-1.8965X_nX_{n-2}^2}  \\
&&\scriptstyle \mathrm{+5.4255X_nX_{n-1}X_{n-2}+0.7258X_{n-1}X_{n-3}^2} \nonumber \\
&&\scriptstyle \mathrm{-1.7684X_{n-2}X_{n-3}^2+1.1800X_{n-2}^2X_{n-3}, \nonumber} 
\end{eqnarray}

These interval extensions were simulated using the initial condition $X_{n-p} = 1$ for $p = 1, 2, 3$.

\begin{figure}[ht!]
\centering
\subfigure[Pseudo-orbits obtained from Python.]{
\includegraphics[width=8cm,height=6cm]{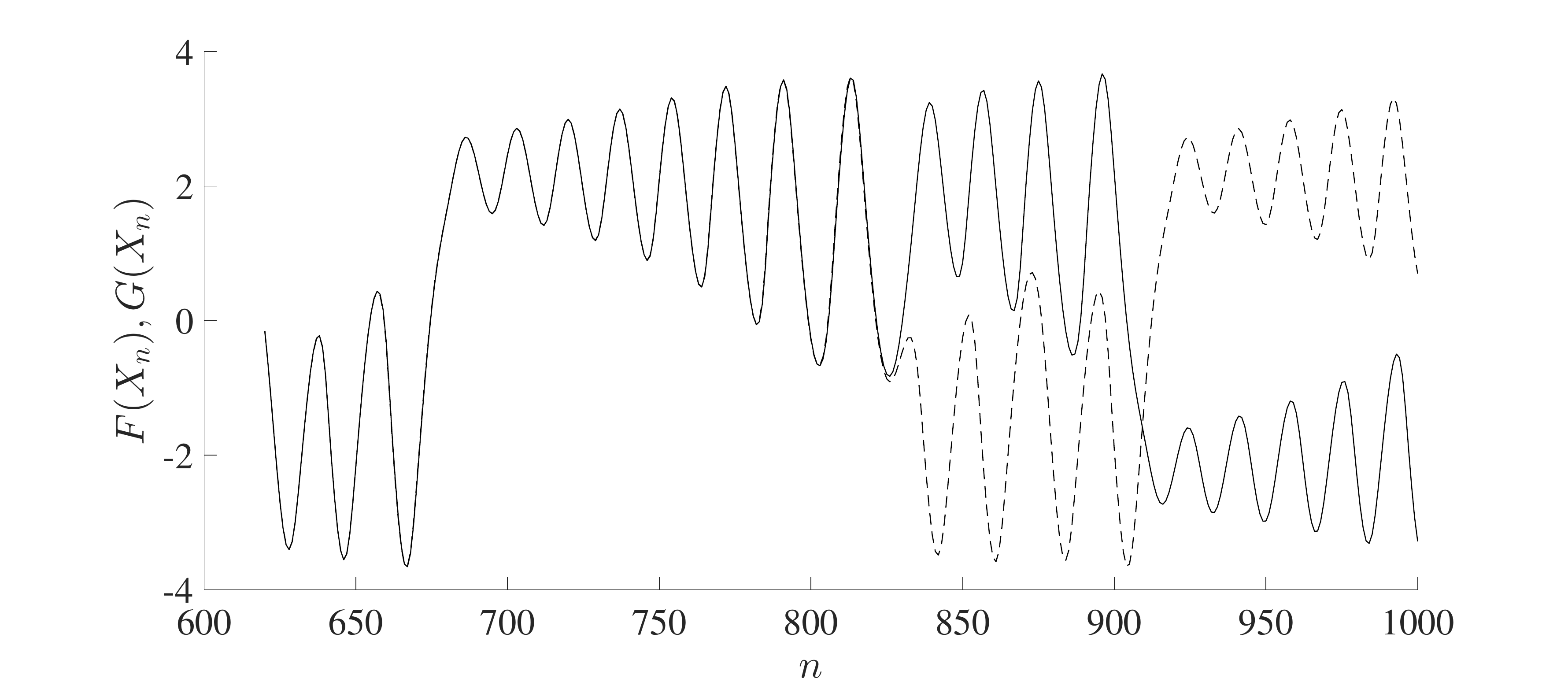}} \label{fig:5a}
\subfigure[Pseudo-orbits obtained from Matlab]{
\includegraphics[width=8cm,height=6cm]{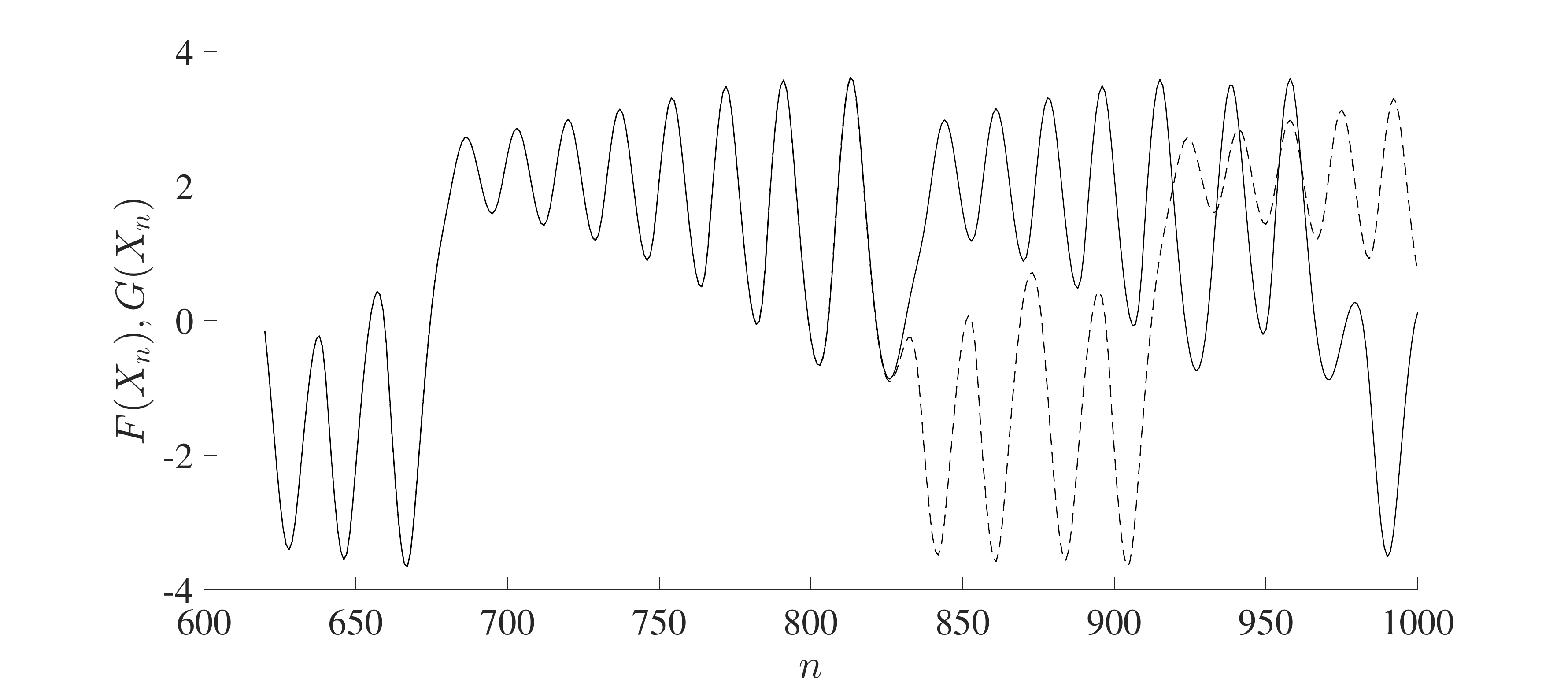}} \label{fig:5b}
\subfigure[Pseudo-orbits obtained from Octave]{
\includegraphics[width=8cm,height=6cm]{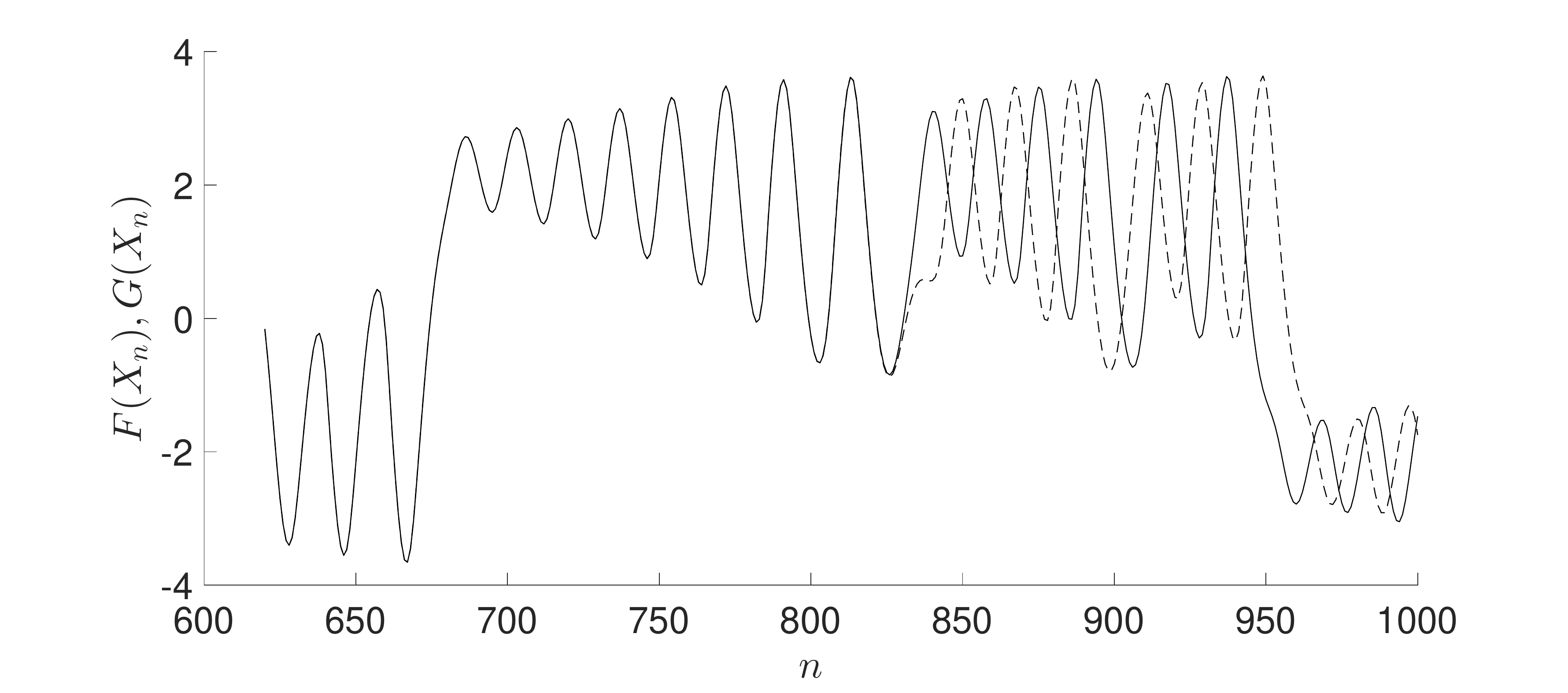}} \label{fig:5c}
\caption{ Chua's circuit: free-run simulation for the interval extensions of Equations \eqref{eq:chua1} and \eqref{eq:chua2}, with results for $F(X_n) (-)$ and $G(X_n) (--)$ and $n$ stands for the number of iterations.}
\label{fig:5}
\end{figure}

\begin{figure}[ht!]
\centering
\subfigure[LBE obtained from Python.]{
\includegraphics[width=8cm,height=6cm]{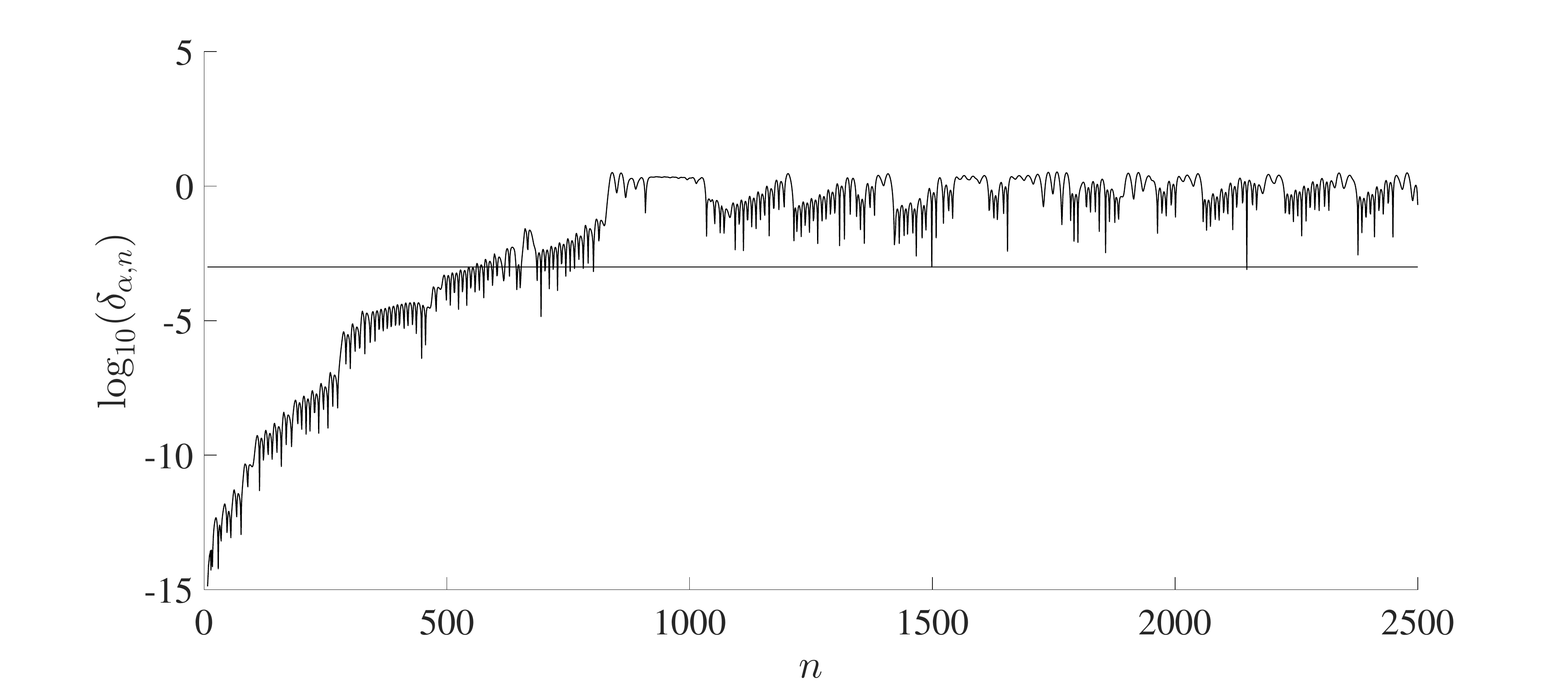}} \label{fig:6a}
\subfigure[LBE obtained from Matlab]{
\includegraphics[width=8cm,height=6cm]{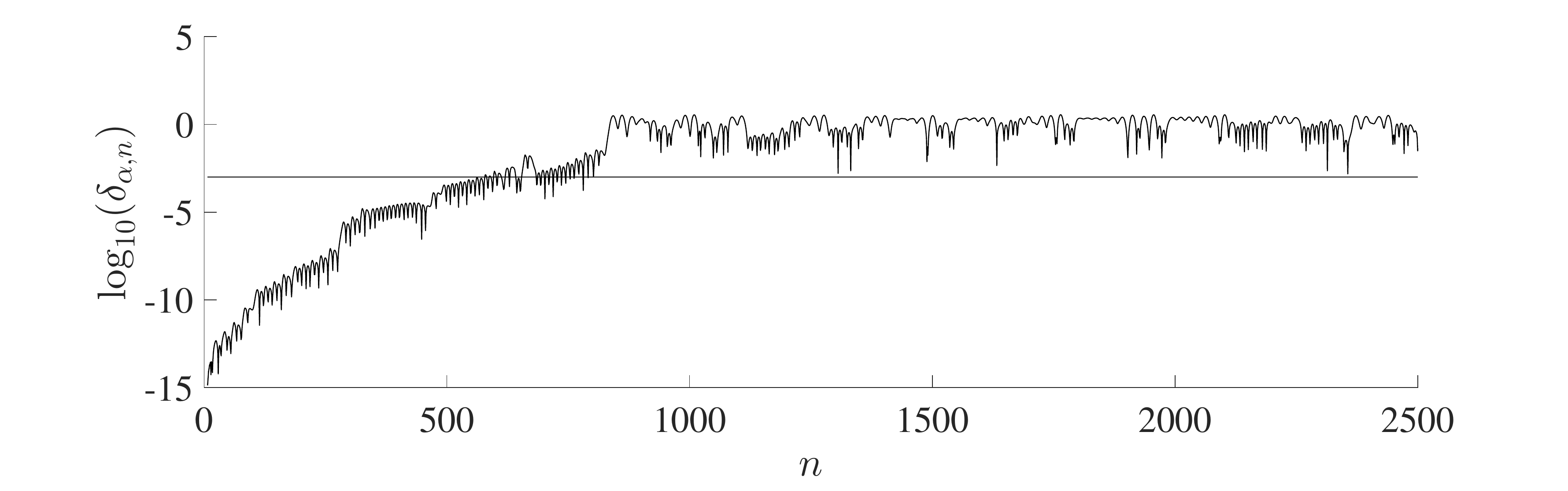}} \label{fig:6b}
\subfigure[LBE obtained from Octave]{
\includegraphics[width=8cm,height=6cm]{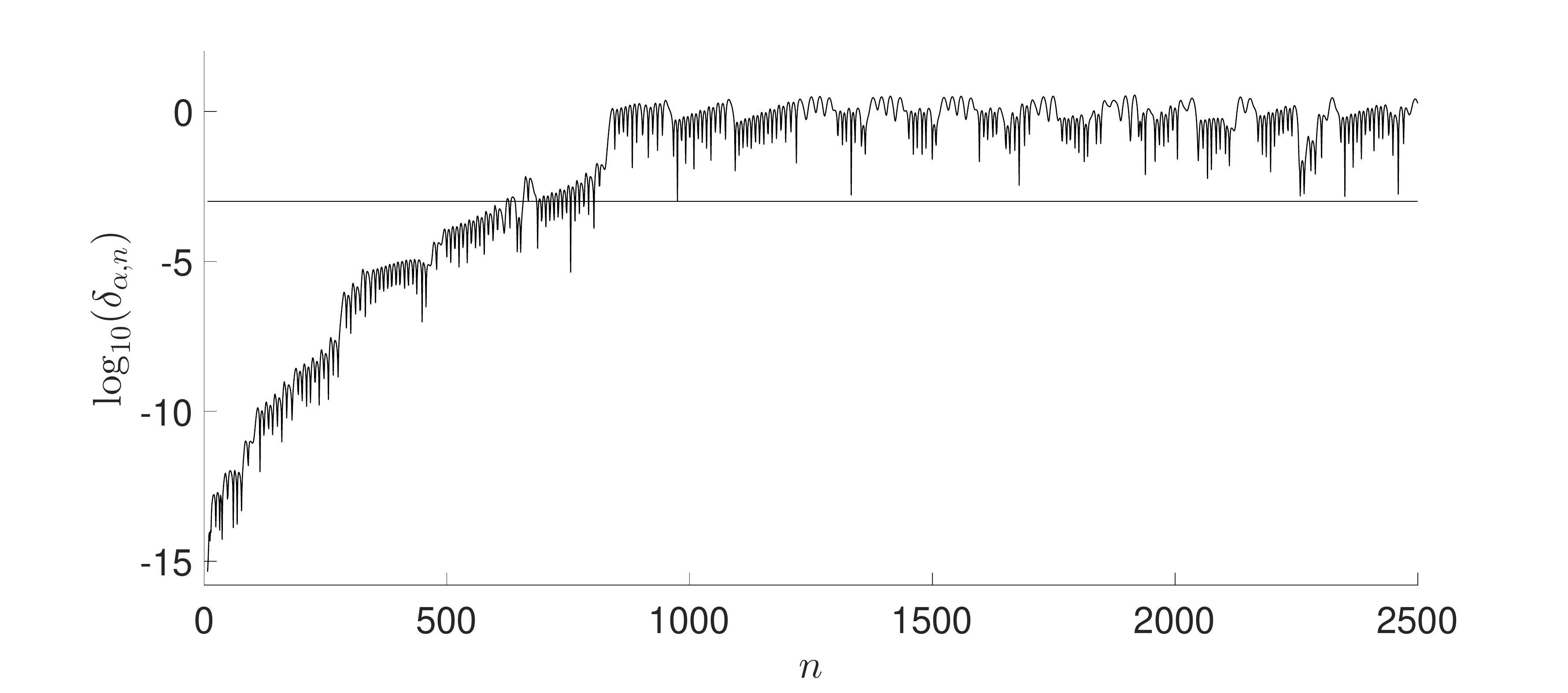}} \label{fig:6c}
\caption{ Chua's circuit: Evolution of lower bound error $\ell_{\Omega,n}$. The values are plotted using $log_{10}$.}
\label{fig:6}
\end{figure}

Figure \ref{fig:5} shows the free-run simulation for the system. As well as the Duffing-Ueda oscillator, for Chua's circuit, running different software cause a different orbit in the time series. Figure \ref{fig:6} shows the evolution of the Lower Bound Error. Analyzing the number of iterations that satisfies $\varepsilon = 0.001$, the simulation is no longer reliable when $n \ge 545$
for Python, $n \ge 579$ for Matlab and $n \ge 624$ for Octave. For this reason, Matlab represents a $6.24\%$ greater reliability compared to Python and Octave represents a $14.5\%$ greater reliability compared to Python. Table \ref{conf} shows the reliability summary of the systems studied for the three proposed softwares.

\begin{table}[!ht]
\centering
\normalsize
\setlength{\tabcolsep}{10pt} 
\renewcommand{\arraystretch}{1.65}
\caption{Number of iterations that the system is not more reliable according to the simulation criterion given by Eq. \ref{eq:stop}.}
\label{conf}
\begin{small}
\begin{tabular}{cccc} \hline
 & Python & Matlab & Octave \\ \hline
Duffing-Ueda & 4450 & 4471 & 5258 \\ 
Chua's circuit & 545 & 579 & 624 \\ \hline
\end{tabular}
\end{small}
\end{table}

In addition, it was calculated the time each software takes to process each algorithm. This was used as a method of comparison, since other works \cite{CNG+2006,Unpingco2008} also use this method for comparison. The algorithm developed in Python was realized with the aid of the NumPy library. And, in each software was simulated a hundred times each algorithm and averaged the time of that hundred times. In addition, the standard deviation was calculated, which indicates how far the data are from the mean.  Table \ref{tab} shows the results found.
\vspace{0.5cm}

\begin{table}[!ht]
\centering
\normalsize
\setlength{\tabcolsep}{10pt} 
\renewcommand{\arraystretch}{1.65}
\caption{Average time of one hundred attempts to execute the proposed algorithm. We have also presented one standard deviation in order to consider the intrinsic fluctuation of time consumption in a computer.}
\label{tab}
\begin{small}
\begin{tabular}{cccc} \hline
 & Duffing-Ueda & Chua's circuit \\ \hline
Matlab & $0.0425 \pm 0.0179$ & $0.0249 \pm 0.0104$ \\ 
Python & $1.4719 \pm 0.266$ & $0.3064 \pm 0.0442$ \\ 
Octave & $3.2808 \pm 0.5919$ & $2.2725 \pm 0.5214$ \\ \hline
\end{tabular}
\end{small}
\end{table}

From Table \ref{tab} it is possible to notice that the software Matlab presents a runtime for the same algorithm much smaller than the time presented by Python. And although the Octave presents a greater reliability for the system, the processing time for this software was much higher.
 
 For the Duffing-Ueda oscillator, the processing time of Python is about $50$ times greater and Octave about $95$ times greater when both are compared with Matlab. Already for Chua's circuit, this time is about $15$ and $152$ times greater for Python and Octave, respectively.

\section{Conclusion}

This work has presented a comparison of the performance of Matlab, Octave and Python  in the simulation of dynamical systems. We have adopted the lower bound error as an index. We have also considered the time cost of the simulation. The LBE offer a way to estimate the maximum number of iterations, or a stop criterion, from which the simulation did not present confidence.

Matlab and Octave work with the functions built into the software itself, while in Python it's called via front-end, as with NumPy. Due to these factors Matlab is significantly faster for the calculation of the proposed algorithm. But, Octave presented a slightly  higher simulation reliability for this application. The Matlab processing time for this application was the shortest. Python, on the other hand, has been the slowest and it has been intermediate regarding the maximum number of iterations. For future works, we have planned to compare these results using the Ocean Code initiative of IEEE \cite{ElHawary2018}.


\bibliography{ifacconf}           
\end{document}